\documentstyle[epsfig]{mn}
\newif\ifAMStwofonts
\ifoldfss
  \ifCUPmtlplainloaded \else
    \NewTextAlphabet{textbfit} {cmbxti10} {}
    \NewTextAlphabet{textbfss} {cmssbx10} {}
    \NewMathAlphabet{mathbfit} {cmbxti10} {} % for math mode
    \NewMathAlphabet{mathbfss} {cmssbx10} {} %  "   "    "
  \fi
  \ifAMStwofonts
    \ifCUPmtlplainloaded \else
      \NewSymbolFont{upmath} {eurm10}
      \NewSymbolFont{AMSa} {msam10}
      \NewMathSymbol{\upi}     {0}{upmath}{19}
      \NewMathSymbol{\umu}     {0}{upmath}{16}
      \NewMathSymbol{\upartial}{0}{upmath}{40}
      \NewMathSymbol{\leqslant}{3}{AMSa}{36}
      \NewMathSymbol{\geqslant}{3}{AMSa}{3E}

    \fi
  \fi
\fi % End of OFSS

\ifnfssone
  \newmathalphabet{\mathit}
  \addtoversion{normal}{\mathit}{cmr}{m}{it}
  \addtoversion{bold}{\mathit}{cmr}{bx}{it}
  \newmathalphabet{\mathbfit} % math mode version of \textbfit{..}
  \addtoversion{normal}{\mathbfit}{cmr}{bx}{it}
  \addtoversion{bold}{\mathbfit}{cmr}{bx}{it}
  \newmathalphabet{\mathbfss} % math mode version of \textbfss{..}
  \addtoversion{normal}{\mathbfss}{cmss}{bx}{n}
  \addtoversion{bold}{\mathbfss}{cmss}{bx}{n}
  \ifAMStwofonts
    \ifCUPmtlplainloaded \else
      \UseAMStwoboldmath
      \makeatletter
      \new@mathgroup\upmath@group
      \define@mathgroup\mv@normal\upmath@group{eur}{m}{n}
      \define@mathgroup\mv@bold\upmath@group{eur}{b}{n}
      \edef\UPM{\hexnumber\upmath@group}
      \new@mathgroup\amsa@group
      \define@mathgroup\mv@normal\amsa@group{msa}{m}{n}
      \define@mathgroup\mv@bold\amsa@group{msa}{m}{n}
      \edef\AMSa{\hexnumber\amsa@group}
      \makeatother
      \mathchardef\upi="0\UPM19
      \mathchardef\umu="0\UPM16
      \mathchardef\upartial="0\UPM40
      \mathchardef\leqslant="3\AMSa36
      \mathchardef\geqslant="3\AMSa3E
    \fi
  \fi
\fi % End of NFSS release 1

\ifnfsstwo
  \DeclareMathAlphabet{\mathbfit}{OT1}{cmr}{bx}{it}
  \SetMathAlphabet\mathbfit{bold}{OT1}{cmr}{bx}{it}
  \DeclareMathAlphabet{\mathbfss}{OT1}{cmss}{bx}{n}
  \SetMathAlphabet\mathbfss{bold}{OT1}{cmss}{bx}{n}
  \ifAMStwofonts
    \ifCUPmtlplainloaded \else
      \DeclareSymbolFont{UPM}{U}{eur}{m}{n}
      \SetSymbolFont{UPM}{bold}{U}{eur}{b}{n}
      \DeclareSymbolFont{AMSa}{U}{msa}{m}{n}
      \DeclareMathSymbol{\upi}{0}{UPM}{"19}
      \DeclareMathSymbol{\umu}{0}{UPM}{"16}
      \DeclareMathSymbol{\upartial}{0}{UPM}{"40}
      \DeclareMathSymbol{\leqslant}{3}{AMSa}{"36}
      \DeclareMathSymbol{\geqslant}{3}{AMSa}{"3E}
    \fi
  \fi
\fi % End of NFSS release 2

\ifCUPmtlplainloaded \else
  \ifAMStwofonts \else % If no AMS fonts
    \def\upi{\pi}
    \def\umu{\mu}
    \def\upartial{\partial}
  \fi
\fi

\title[Thermal and Non-thermal Plasmas in the Galaxy Cluster~3C~129]
  {Thermal and Non-thermal Plasmas in the Galaxy Cluster~3C~129}
\author[Krawczynski et al.]
  {H.~Krawczynski$^{1}$,
   D.E.~Harris$^2$,
   R.~Grossman$^3$,
   W.~Lane$^4$,
   N.~Kassim$^4$,
   A.G.~Willis$^5$\\
  $^1$  Washington University in St. Louis, Phys.\ Dept.,
 	1 Brookings Drive, Campus Box 1105, St. Louis, MO 63130, USA \\
  $^2$ Smithsonian Astrophysical Observatory, 60 Garden Str., 
       Cambridge, MA 02138, USA\\
  $^3$ Department of Physics and Astronomy, Barnard College, Columbia
       University, New York, NY 10027, USA\\
  $^4$ Code 7213, Remote Sensing Division, Naval Research Laboratory, 
       Washington, DC 20375-5351, USA\\
  $^5$ National Research Council of Canada, Herzberg Institute of
       Astrophysics, Dominion Radio Astrophysical Observatory,\\
       P.O. Box 248, Penticton, BC, Canada V2A 6K3}
\date{Accepted --- --. Received 2002, Aug 20}
\pubyear{2002}

\begin{document}
\label{firstpage}
\maketitle
\begin{abstract}
We describe new Chandra spectroscopy data of the cluster which harbors the 
prototypical ``head tail'' radio galaxy 3C~129 and the weaker radio galaxy 
3C~129.1. We combined the Chandra
data with Very Large Array (VLA) radio data taken at 0.33, 5, and
8~GHz (archival data) and 1.4~GHz (new data). We also obtained new HI
observations at the Dominion Radio Astrophysical Observatory (DRAO) to
measure the neutral Hydrogen column density in the direction of the
cluster with arcminute angular resolution.
The Chandra observation reveals extended X-ray emission from the radio galaxy
3C 129.1 with a total luminosity of 1.5$\times 10^{41}$ erg s$^{-1}$.
The X-ray excess is resolved into an extended central source of
$\simeq$2\arcsec (1 kpc) diameter and several point sources
with individual luminosities up to 2.1$\times 10^{40}$ erg s$^{-1}$.
There is no evidence for a correlation between 
the 3C 129.1 X-ray and radio morphology.
In the case of the radio galaxy 3C~129, the Chandra observation shows,
in addition to core and jet X-ray emission reported in an earlier paper,
some evidence for extended, diffuse X-ray emission from a region east 
of the radio core. The 12\arcsec$\times$36\arcsec (6 kpc $\times$ 17 kpc) 
region lies ``in front'' of the radio core, in the same direction 
into which the radio galaxy is moving.
We use the radio and X-ray data to study in detail the pressure balance 
between the non-thermal radio plasma and the thermal Intra Cluster Medium (ICM)
along the tail of 3C~129 which extends over 15\arcmin (427 kpc).
Depending on the assumed lower energy cutoff of the electron energy spectrum,
the minimum pressure of the radio plasma lies a factor of between 
10 and 40 below the ICM pressure for a large part of the tail. 
We discuss several possibilities to explain the apparent pressure mismatch.
\end{abstract}
\section{Introduction}
\subsection*{Observations of Radio Galaxies within Galaxy Clusters}
Tailed radio galaxies within galaxy clusters have drawn
intensive study for several decades on both theoretical
and observational grounds.
Radio and X-ray data complement each other in that the radio data show 
high resolution images of the jets and more extended synchrotron features 
yielding information on strength and orientation of the magnetic
field within the radio plasma and between the radio galaxy and the
observer, while the X-ray data gives information about the temperature, 
chemical composition, density, and  pressure of the Intracluster Medium (ICM).  
Thus, the jets propagate into a well defined medium, which
greatly facilitates the interpretation of the radio images. 
It becomes possible to study, for example, the pressure balance between
the radio-plasma and ICM, and thus to search for hidden pressure components
in the radio plasma. 
Various authors have studied the radio plasma/ICM pressure
balance for FR~I and FR~II radio galaxies, in most cases
finding the minimum synchrotron pressure lower than the
ICM pressure by approximately one order of magnitude
(e.g.\ Feretti et al.\ 1990, Hardcastle \& Worrall 2000 and
references therein).
Furthermore, the long tails of radio galaxies carry the imprint of 
relative motion between the non-thermal plasma and the ambient hot gas. 
In the parlance of the field, they reflect the weather conditions in the ICM. 
The potential of such observations to reveal details of cluster mergers 
such as subsonic and transonic bulk flows, shocks, and turbulence has been 
investigated on theoretical grounds by a number of authors 
(Loken et al.\ 1995, Burns et al.\ 2002).
On the observational side, G\'{o}mez at al.\ (1997) studied a large 
number of Wide Angle Tail (WAT) 
radio galaxies inside Abell clusters. They found a strong correlation between 
the orientation of the radio tails and the elongations of the ICM surface 
brightness profiles, and interpreted it as evidence for bulk plasma motion 
that follows cluster mergers. 
Bending of radio tails has often been explained by invoking motion 
of the radio galaxy in combination with ``stormy ICM weather'' 
(e.g.\ Feretti et al.\ 1985, Sakelliou, Merrifield and McHardy 1996)
but this interpretation does not work for all bent radio tails (e.g.\ 
Eilek et al.\ 1984).

In addition to facilitating studies of radio galaxies and the ICM themselves,
observations of radio galaxies in galaxy clusters make it possible to study 
the interaction of the radio galaxies with the ICM and vice versa.
The radio plasma can interact with the ICM and form so-called
cavities devoid of ICM that show up as depressions in the X-ray surface 
brightness maps. 
Cavities have been observed in the Perseus cluster B\"ohringer et al.\ (1993), 
in the Cygnus A cluster by Carilli et al.\ (1994); more recently, CHANDRA 
observations revealed cavities in the clusters Hydra A, A~2597, A~2052,
and A~4059 
(McNamara et al.\ 2000, 2001, Blanton et al. 2001b, Heinz et al.\ 2002). 
It is not yet clear whether these cavities are merely curiosities 
or, whether they influence the ICM in an important way by supplying 
a dynamically important amount of heat and/or magnetic field to the ICM
(Br\"uggen et al. 2002, Quilis, Bower, \& Balogh 2001, 
Reynolds, Heinz, \& Begelman 2002, Fabian et al.\ 2002).
The relative motion of the ICM and the Interstellar Medium (ISM) of the
radio galaxy's host galaxy can result in a modification of the 
ISM and in a bow shock forming ``in front'' of the
radio galaxy (e.g.\ Stevens, Acreman, Ponman 1999,
Toniazzo \& Schindler 2001, and references therein). 
Although these interactions can also be observed in normal galaxies,
the radio maps of radio galaxies can be used to better interpret the data by
constraining the motion of the galaxies in the plane of the sky 
as well as the ICM magnetic field.
\subsection*{The Cluster 3C~129 and Outline of the Paper}
In this paper we present Chandra data on the nearby galaxy cluster 3C~129 
($z\,=$ 0.0223).
Owing to its low Galactic latitude ($l$\,= 160$^\circ$.5, $b$\,= 0.3$^\circ$) 
it has not been studied extensively at optical wavelengths.  In
addition to velocity measurements of 3C~129 and 3C~129.1 (Spinrad,
1975), we are aware of only one other velocity-verified cluster galaxy: 
WEIN 048 (Nilsson et al.\ 2000).
Optically, the galaxy 3C~129 has been described as a
weak elliptical galaxy without peculiar features \cite{Coli:90}.
X-ray observations with the Einstein, EXOSAT, and ROSAT satellites 
\cite{Edge:91,Leah:00} yielded a cluster luminosity of 2.7$\times 10^{44}$~ergs s$^{-1}$ 
in the 0.2-10~keV energy band, a central gas density of 6$\times 10^{-3}~\rm cm^{-3}$ 
and an ICM mass of $3.6\times10^{13}$ $\rm M_\odot$. 
Based on Einstein IPC data Leahy \& Yin (2000) found evidence for a cooling 
flow with $\dot{\rm M}\approx85~ \rm M_\odot~ \mbox{yr}^{-1}$. 
However, reanalyzing the same data, Taylor et al.\ (2001) derived an 
upper limit of $\dot{\rm M}\,<$ 1.2 M$_\odot \mbox{yr}^{-1}$.
Based on the Chandra imaging spectroscopy data 
we performed a spectroscopic deprojection analysis, 
taking into account the asymmetry of 3C~129's X-ray surface brightness by 
using a set of elliptical ICM shells
(Krawczynski 2002, called Paper I in the following). 
This analysis did not show a decrease of the ICM temperature toward the cluster 
center, as expected for a cooling flow.

The cluster contains two radio galaxies. The radio source 3C~129 is
a well studied galaxy in the radio wavelength range as
a prototype head-tailed (HT) galaxy with a strongly curved, 
15\arcmin~long (426~kpc), two armed tail \cite{Mile:73,VanB:82,Kass:93,Fere:98}. 
In an earlier paper (Harris, Krawczynski \& Taylor, 2002, called Paper II in the following) 
we presented evidence for X-ray emission from the inner northern jet of 3C~129.
Near the projected center of the X-ray galaxy cluster is the weaker and
smaller radio source 3C~129.1. 
This galaxy also has dual radio jets which extend over $\simeq$2~arcmin 
(57 kpc) \cite{Down:84,Jaeg:87,Kass:93}.

In this paper we describe X-ray and radio observations of this cluster. 
%
% Contents of paper
%
We complement the Chandra X-ray data with archival 0.33~GHz, 5~GHz and 8~GHz and 
with new 1.4~GHz Very Large Array (VLA)\footnote{The
National Radio Astronomy Observatory is a facility of the National
Science Foundation operated under cooperative agreement by Associated
Universities, Inc.} radio data. 
We use the radio and X-ray data to search for evidence of interactions
between the radio plasma and the ICM and to study the pressure balance between
the two plasmas.
Since the cluster lies at low Galactic latitude we expect substantial 
variations of the neutral Hydrogen column density N$_{\rm H}$ over the 
field of view of the Chandra observations. We therefore obtained 
HI observations with the DRAO telescope.
Tables \ref{data1} and \ref{data2} summarize the data sets used in our 
analysis.

The remainder of the paper is structured as follows.
We introduce observations and data sets in Sect.~\ref{pdata}.
Results of radio observations are presented in Sect.\ \ref{radio},
while in Sect.~\ref{cg} we discuss radial profiles of the ICM surface brightness 
distribution together with the geometry of the ICM.
We describe the search of the X-ray data for 
signatures of radio galaxy/ICM interactions in Sect.~\ref{features}, and 
examine the pressure balance between the plasma of 3C~129 
and the ICM in Sect.~\ref{pb}. Finally, we summarize our results
in Sect.~\ref{disc}.

We use $H_0=$ $h_0$ 100~km~s$^{-1}$~Mpc$^{-1}$ with $h_0\,=0.65$ throughout, 
setting the cluster at a distance of 100~Mpc; 1 arcsec corresponds to 474~pc.
We quote all errors on 90\% confidence level.
All coordinates are given for equinox 2000. Distances from the radio core
of 3C~129 are given as point to point distances, and not as distances along 
the radio tail. 
\section{Data Sets and Data Analysis}
\label{pdata}
\subsection*{HI Observations with the Dominion Radio Astrophysical Observatory}
Since the 3C~129 area lies close to the Galactic plane it is included in the 
21~cm line and continuum observations planned as part of the second phase of the
Canadian Galactic Plane Survey (CGPS; see Taylor 1999 for an overview of its goals).
The DRAO Synthesis telescope HI observations of this part of the sky 
were completed early in the second phase of the survey. They were reduced following 
the standard HI reduction procedure for the CGPS, described by Higgs (1999). 
This work includes the insertion of ``short spacing data'' collected with the 
DRAO 26~m telescope and corrected for stray radiation effects. 
So the HI data are fully sampled out to an interferometer spacing of 
approximately 2900 wavelengths.  The resulting images have a resolution of 
1~arcmin (Full Width Half Maximum, FWHM) 
in right ascension and 1~arcmin cosec($\delta$) in declination $\delta$.
We used a method similar to that described by Normandeau (1999), to
determine the optical depth in the direction toward the 3C~129 cluster.
Briefly, HI spectra in the direction of the heads of the radio galaxies 
3C~129 and 3C~129.1, which show absorption, were compared with spectra
toward nearby ``off source'' positions. The comparisons indicate that the
optical depth along the line of sight to these two strong sources
averages to less than 1. We assume that this optically thin condition
holds anywhere in the direction toward the 3C~129 cluster, so we can
derive the HI column density from the expression:
\begin{equation}
N_{\rm H}\,\,=\,\,1.823 \times 10^{18} \times \Sigma_i T_{\rm b}(v_i) \Delta 
{v_i}
\end{equation}
Here, N$_{\rm H}$ is given in atoms~cm$^{-2}$, the sum runs over all
velocities $v_i$ which contain Galactic hydrogen, $T_{\rm b}$ is the
radio brightness temperature in degree Kelvin, and $\Delta v_i$ is the
width of the $i$th velocity interval in km s$^{-1}$.
\subsection*{Radio Observations with the Very Large Array}
The 330~MHz VLA data were acquired during 175 snapshot observations made
over several years in the late 1990's. The data were taken in A, B,
and C configurations.  The final image has $\approx$~8\arcsec resolution (FWHM).
The Root Mean Square (RMS) noise in the map is $\sim 0.75$~mJy beam$^{-1}$, and
the Dynamic Range (DR, the signal-to-noise ratio of the peak brightness)
is $\sim 400$.  A detailed description of these observations has been
given by Lane et al.\ (2002).
 
Data for the 1400 MHz image were obtained at the VLA on August 16, 2001 
(4 hrs, C-configuration), and December 6, 2001 (1.5 hrs, DnC-Configuration).
Short observations of 3C~48 were used for amplitude
calibration, and the data were combined and mapped with the package 
\verb+AIPS+ (Astronomical Imaging Processing System) using standard imaging techniques. 
The image has a FWHM resolution of $\approx$~18\arcsec $\times$ 14\arcsec 
at a position angle of -15$^o$, an RMS noise of 100 $\mu$Jy beam$^{-1}$, 
and a DR$\sim 3000$.
The primary aim of the VLA observations at L-band was to derive
accurate 0.33 GHz - 1.4 GHz spectral indices.  The 330 MHz data have a
minimum UV-spacing of 40$\lambda$, and a maximum of 40.65k$\lambda$ 
(k stands for kilo); for the 1400 MHz data, the minimum 
spacing is 180$\lambda$ and the maximum is 16.5k$\lambda$.  
To make the spectral index maps, both datasets were limited to a maximum 
UV of 15k$\lambda$ and a Gaussian taper was applied during imaging to match 
the measured beam sizes as closely as possible.  
A circular restoring beam of 18'' x 18'' was used in the final maps.
We estimated the errors on the spectral indices by adding/subtracting 5
sigma to fluxes at one frequency and subtracting/adding 5 sigma from
the other. This method gives conservative 5 sigma errors of 0.01 near the
radio core and 0.25 at the end of the radio tails.

Finally we used several VLA maps of 3C 129 and 3C 129.1 with various effective angular 
resolutions that G.\ Taylor kindly provided.
The pressure determination of the 3C 129 radio plasma uses a 5 - 8 GHz spectral index map
extracted from two data sets with identical angular resolutions of 1.8\arcsec 
(Taylor et al.\ 2001).
The width of the 3C 129 tail was determined at distances from 0\arcsec to 20\arcsec, 
0.3\arcmin to 2.5\arcmin, and 2.5\arcmin to 15\arcmin from the radio core
from the 8 GHz, 0.83\arcsec~angular resolution map from Harris et al.\ (2002),
the 8 GHz, 1.8\arcsec~angular resolution map from Taylor et al.\ (2001), and
the 0.33 GHz, 8\arcsec~angular resolution map from Lane et al.\ (2002), respectively.

A high angular resolution map of 3C 129.1 (0.81\arcsec $\times$ 0.76\arcsec~FWHM) 
has been used for the radio contours shown in Fig.\ \ref{rg1}.
\subsection*{Chandra Observations}

Our data consist of two pointings: a 31.4 ksec exposure taken on
December 9, 2000 with the ACIS~S, 6 chip array (aimpoint close to
3C~129), and a 9.6~ksec exposure with the standard ACIS~I CCD on
January 9, 2001 (aimpoint close to 3C~129.1).
The analysis uses the \verb+CIAO 2.2+ software. 
Filtering for major background flares reduces the ACIS S data set to 
28.3~ksec and the ACIS~I data set to 9.3 ksec.
We produce flat-fielded X-ray brightness maps using a soft (0.5-1~keV), 
an intermediate (1-2 keV) and a hard (2-5~keV) energy band.  
The brightness maps were corrected for instrumental parameters such 
as the effective area etc.\ by dividing by the appropriate exposure maps.
The flat-fielded maps give the flux of 0.5-5 keV photons per $\rm cm^2\,s\,arcsec^2$. 
We combine the flat-fielded maps of both observations by 
weighting each one with the corresponding integration time.  
Further details concerning these data can be found in Paper I.

Some X-ray features have only modest statistical significance.
Using the number of counts in the source ``ON region'' $N_{\rm ON}$,
and the number of counts in the comparison ``OFF region'' $N_{\rm OFF}$,
we compute the statistical significances based on the method of Li \& Ma (1983). 
The method is based on Poisson probability distributions and 
correctly accounts for the statistical uncertainty of the counts
in both the source and comparison regions.
The statistical significance of an excess is given in standard deviations by:
\begin{equation}
\label{lima}
S(N_{\rm ON},N_{\rm OFF};\xi)\,=\,\sqrt{-2\, \ln(\lambda)}
\end{equation}
where
\begin{equation}
\lambda\,=\,
\left[\frac{\xi}{\xi+1}\,
\left(\frac{N_{\rm ON}+N_{\rm OFF}}{N_{\rm ON}}\right)\right]^{N_{\rm ON}}\,\times\,
\end{equation}
\[
\hspace*{1cm}
\left[\frac{1}{\xi+1}\,
\left(\frac{N_{\rm ON}+N_{\rm OFF}}{N_{\rm OFF}}\right)\right]^{N_{\rm OFF}}
\]
and $\xi$ given by:
\begin{equation}
\xi\,=\,
\frac{\int_{\Delta \Omega_{\rm ON}}\,A(\Omega)\,d\Omega}
     {\int_{\Delta \Omega_{\rm OFF}}\,A(\Omega)\,d\Omega}
\end{equation}
where $\Delta \Omega_{\rm ON}$ and $\Delta \Omega_{\rm OFF}$ are the
solid angle areas of the source and background regions, respectively, 
and $A(\Omega)$ is the the effective area of the detector.
We use the exposure 
map produced for 1.7 keV photons as $A(\Omega)$ and 
integrate over this map with the CIAO tool \verb+dmstat+.
Although the method is based on Poisson probability distributions,
the expression for the statistical significance is only valid for
$N_{\rm ON}> 10$ and $N_{\rm OFF}> 10$ \cite{Alex:92}.
If this condition is not met, we use the equation of Alexandreas et al.\ (1992)
\begin{equation}
P_{\rm c}\,\,=\,\,
\sum_{n=N_{\rm ON}}^{\infty}\,\,
\frac{\eta^n}{(1+\eta)^{N_{\rm OFF}+n+1}}\,\,
\frac{(N_{\rm OFF}+n)!}{N_{\rm OFF}!\,n!} 
\end{equation}
to compute the chance probability $P_{\rm c}$ that $N_{\rm ON}$ counts are observed
by chance, taking into account the statistical uncertainty of the ``OFF'' 
measurement.
We state the significances without taking into account ``trial factors''
due to the search for features over an extended solid angle region. 
\section{Radio Maps}
\label{radio}
\begin{figure}
\begin{center}
\hspace*{-0.1cm} \epsfig{file=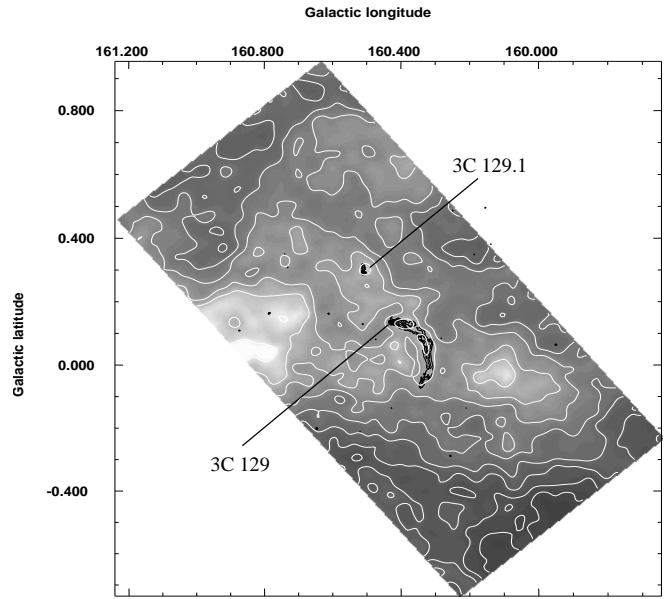,width=3.4in}
\end{center}
\caption{ 
Neutral Hydrogen Column density in the direction of the cluster 3C~129
derived from the DRAO HI data, smoothed with a Gaussian of 2.45\arcmin~FWHM. 
Darker regions show higher N$_{\rm H}$ column densities. 
Rather atypically, the column densities are a bit lower along the
galactic plane than at slightly lower and higher latitudes.
The ten contour levels are linearly spaced from 
8$\times10^{21}$ to $10^{22}$ cm$^{-2}$. Compared to nearby fields 
in the Galactic plane (see e.g.\ Leahy and Roger 1991, Figs.\ 7 and 8), 
only modest gradients of the N$_{\rm H}$-column density are observed.
}
\label{HImap}
\end{figure}

The N$_{\rm H}$-map derived from the HI observations, 
smoothed with a Gaussian of 2.45\arcmin~FWHM is shown in Fig.~\ref{HImap}. 
Although the N$_{\rm H}$-values change 
by 25\% over the 1.25$^\circ$$\times$2.5$^\circ$ field of view, 
the Chandra observations lie in a region of rather constant values of 
between 8.4$\times$$10^{21}$~cm$^{-2}$ and 9.2$\times$$10^{21}$~cm$^{-2}$.
Strasser and Taylor (2002) studied HI absorption spectra of 273 background
sources from the Canadian Galactic Plane Survey. In their Figure 2 they
show that in optically thin regions the apparent off-source temperatures 
cover a range of about 100~K (upper bound) to 50~K (lower bound). 
Assuming a mean value of 75 K, these off-source data sets imply a
maximum error in the HI determination of about 30\%.
Our results are marginally consistent with those of Leahy \& Yin (2000) and 
Taylor et al.\ (2001), derived from fitting the cluster's X-ray energy spectra: 
the first authors report N$_{\rm H}\,=$ (5.7$\pm$0.3)$\times$$10^{21}$~cm$^{-2}$ 
based on ROSAT and EXOSAT data, and the latter authors get N$_{\rm H}\,=$ 
(6.4$\pm$0.4)$\times$$10^{21}$~cm$^{-2}$ from ASCA data.

\begin{figure}
\begin{center}
\epsfig{file=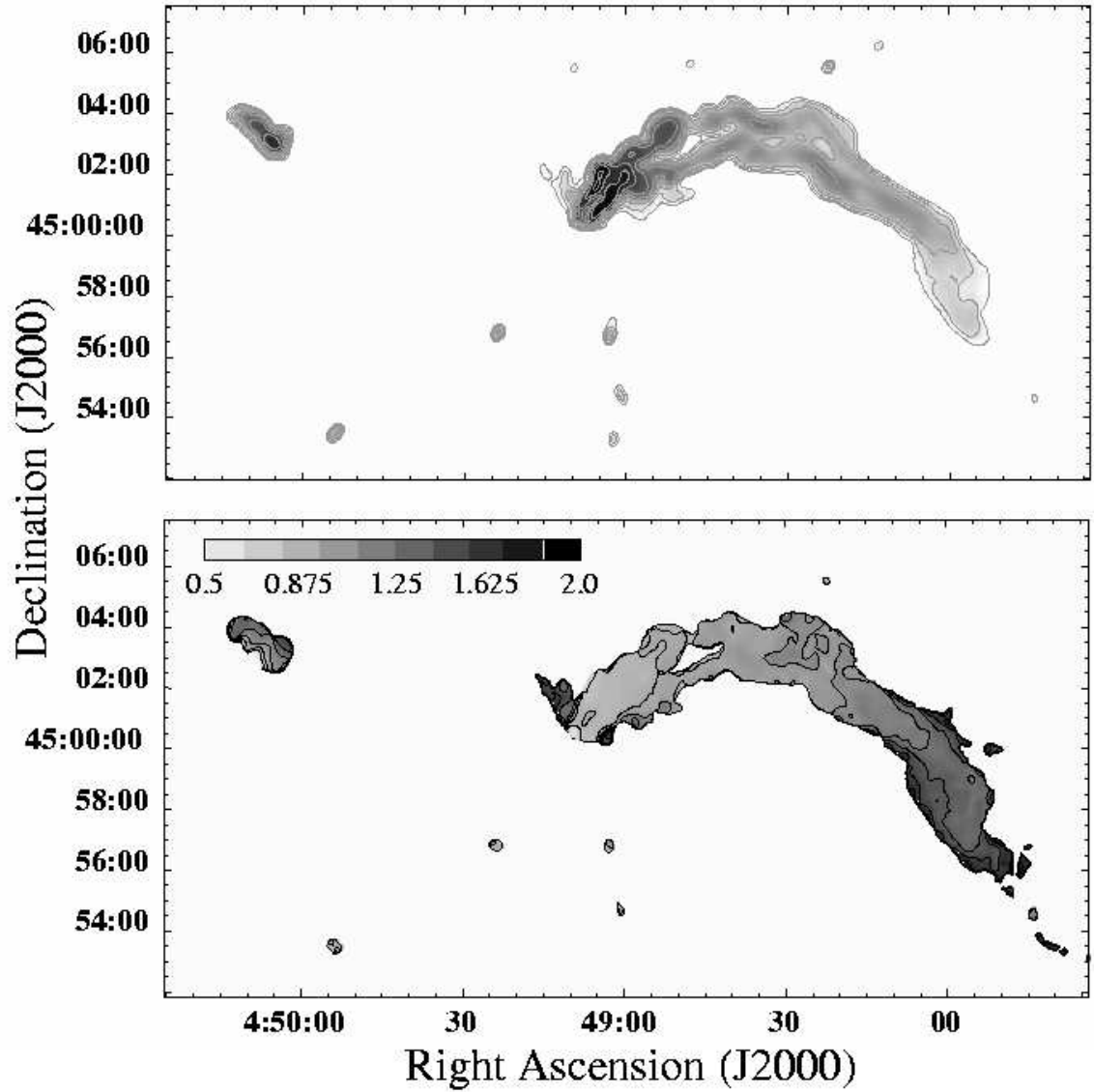,width=6.6in}
\end{center}
\caption{ The upper panel shows the 1.4~GHz surface brightness
distribution.  The peak surface brightness is 0.29 Jy beam$^{-1}$ 
and the 8 contours are logarithmically spaced from 0.001 to 0.256 Jy beam$^{-1}$
(beamwidth 16\arcsec FWHM).
Darker regions correspond to a higher surface brightness.
The lower panel shows the 0.33-1.4 GHz spectral indices ($\alpha$ from
$S_{\nu}\,\propto$ $\nu^{-\alpha}$) derived from the 0.33 GHz data of
Lane et al.\ (2002) and the 1.4 GHz data of the upper panel.  Darker
regions show relatively steeper energy spectra (larger $\alpha$) and the
7 contour levels are linearly spaced from $\alpha\,=$ 0.5 to
$\alpha\,=$ 2. ($\Delta \alpha\,=$ 0.25).  The small radio source in
the east is 3C~129.1 and the larger one in the west is 3C~129.}
\label{w1}
\end{figure}

The 1.4 GHz radio surface brightness distribution of the 3C~129.1 and
3C~129 region is shown in Fig.~\ref{w1}, upper panel. The map shows
substantially more detail than the previously published map at
this frequency \cite{Jaeg:87}.  
One clearly recognizes the peculiar radio feature 
(called ``radio crosspiece'' in the following) 
to the northeast of the radio head seen in earlier low-frequency 0.33 GHz data
\cite{Jaeg:83,Lane:02}.  Figure \ref{w1}, lower panel, shows the
0.33~GHz-1.4~GHz spectral indices ($\alpha$ defined by $S_{\nu}\,\propto$
$\nu^{-\alpha}$) derived from the 0.33 GHz data of Lane et al.\ (2002)
and our 1.4 GHz data.  The spectrum is flattest at the 3C~129 radio
core ($\alpha_{\rm core}\,=$ 0.5).  Between 0\arcmin~and 6\arcmin~from the radio
core the spectrum of both radio tails stays rather constant with
spectral indices from 0.5 to 0.8; between 6\arcmin~and 11\arcmin~from the radio
core it steepens to $\alpha\,\approx$ 1; and from 11\arcmin~to 15\arcmin~it varies
from 1 to 1.5. The region of the radio crosspiece has spectral indices of
between 1.1 and 1.5. A natural interpretation of the spectral index
distribution in the tail is that the electron population
has an energy spectrum ($dN/dE\,\propto$ $E^{-p}$) with $p\,\approx$
2.0.  Within a distance of 8\arcmin~from the radio core the radio spectrum
steepens from $\alpha\,=$ 0.5 to 1.0, and thus the electron spectrum
steepens from $p\,=$ 2 to 3, as a consequence of synchrotron 
and Inverse Compton cooling of the radio plasma. Further cooling eventually 
results in $\alpha>$1, once the electron spectra have cooled enough so that 
the high energy cutoff of electrons is low enough to affect the frequency 
band being observed. Weakening magnetic fields further down the radio tail
will also contribute to this effect.

Assuming no re-acceleration of electrons in the radio tail 
and using an order of magnitude magnetic field strength of $B\,=$ 5$\mu$G, 
we estimate the age of the electrons at 8' from the radio core where
the spectral index of the synchrotron radiation is 0.5.
For this purpose, we assume an initial electron energy spectrum 
$dN/dE\,\propto$ $E^{-p}$ with $p\,=$ 2.0, and compute the 0.33~GHz-1.4~GHz 
spectral index taking into account synchrotron and Inverse Compton
energy losses of the electrons (see Myers \& Spangler, 1985).
Following Jaffe \& Perola (1974) we assume that electrons are isotropized 
on time scales short compared with their radiative lifetimes.
In this approximation, Inverse Compton cooling on the Cosmic Microwave 
Background can be taken into account by using an effective magnetic field
$B'\,=$ $\sqrt{B^2+B_{\rm CMB}^2}$ with 
$B_{\rm CMB}\,=$ $3.25\,(1+z)^2$~$\mu$G 
to compute the energy losses of the electrons. 
The model reproduces the observed spectral index $\alpha\,=$ 1 for an age of 
95~Myr of the electrons. This age translates into a velocity of the 
galaxy 3C~129 of:
\begin{equation}
v_{\rm gal}\,\approx 2260 \sin{(\theta)}^{-1}
(h_0/0.65)^{-1}\,\mbox{km~s}^{-1}
\label{vgal}
\end{equation}
where $\theta$ is the angle between the plane of motion of 3C~129 and the line 
of sight toward the observer. 
Although the equipartition magnetic field at 8\arcmin~from the radio core 
derived from the radio data is smaller (between 1.4 and 3.7 $\mu$G
for lower frequency cutoffs $\nu_1\,=$ 330 MHz and $\nu_1\,=$ 1 MHz, 
respectively) the radio plasma - ICM pressure balance argues for a stronger 
magnetic field, possibly up to 15 $\mu$G. 
Since $v_{\rm gal}$ depends strongly on $B$, a range of allowed 
$v_{\rm gal}$-values tightly constrains the magnetic field (see Sect.\ \ref{disc}).

Lane et al.\ (2002) interpreted the radio crosspiece as fossil non-thermal 
electrons re-energized by the bow-shock of the radio galaxy 3C~129, supersonically
traversing the ICM. Assuming that the segment of the energy spectrum
of the electrons responsible for the 0.33 to 1.4 GHz emission in the radio 
crosspiece has not yet steepened due to radiative cooling, the spectral 
index $\alpha\,\approx$ 1.3 translates into a particle spectral index 
of $p\,=$ $3.6$. 
Invoking the standard theory of diffusive particle acceleration at shocks
\cite{Drur:83}, 
we derive an estimate of the Mach number of $M_{\rm gal}\,=$ 1.87.
For an ICM temperature of 5~keV, the ICM sound speed is 1160 km s$^{-1}$, 
and $M_{\rm gal}\,=$ 1.87 corresponds to $v_{\rm gal}\,=$ 2170 km s$^{-1}$ in fair
agreement with above estimate.
The difference of the radial velocities of 3C~129.1 ($z\,=$ 0.022265) and 3C~129 
($z\,=$ 0.020814) is only 435 km s$^{-1}$, suggesting that the galaxy's velocity 
vector lies mainly in the plane of the sky. 

The spectral index values and variations found across the source between data at 
0.33 and 1.4 GHz are in reasonable agreement with those found between 
0.612 and 1.4 GHz (van Breugels 1982; J\"{a}gers 1987), and between 
1.4 and 4.1 GHz (Miley 1973). 
Comparing our results with the 0.074 - 0.61 GHz spectral indices of Perley \& Erickson (1979), 
we find that the spectrum flattens at low frequencies, while comparisons 
with the 0.408-2.7 GHz spectral indices of Riley (1973) and the 2.7-4.8 GHz 
and 4.8-10.6 GHz spectral indices of Feretti et al.\ (1998) show that the 
energy spectrum steepens at higher frequencies.
\section{The Geometry of the Intracluster Medium and Discrete Sources}
\label{cg}
The upper panel of Fig.\ \ref{smooth} shows the flat-fielded data of
the two Chandra observations, summed with a weight proportional to the
integration time and smoothed with a Gaussian of 18.8\arcsec FWHM.
Neither before nor after smoothing the X-ray surface brightness do 
we detect obvious sharp gradients, like those found for a large number of
more regular clusters (Markevitch et al.\ 2002) which might be produced 
by sloshing of the central ICM. The cluster has a considerable ellipticity. 
Using surface brightness iso-contours we derive ellipticities 
$\epsilon\,=$ $(a-b)/a$ of about 0.3, $a$ being the semi-major axis, and $b$ the 
semi-minor axis.
% 1.08 vs.: 0.75 
%
The surface brightness iso-contour ellipses are roughly aligned in
the east-west direction which depends slightly on the diameter of the
ellipses.  The surface brightness peaks at and around the location of two
discrete sources: an extended source near the core of the radio galaxy 3C~129.1
($\alpha\,=$ 04$^{\rm h}$50$^{\rm m}$06$^{\rm s}$.6308 and $\delta\,=$
45$^\circ03^\prime05.990^{\prime\prime}$), and a point source,
called FS1 in the following, at 2.0 arcmin toward the southwest of the center.
We use the location $\alpha\,=$ 04$^{\rm h}$50$^{\rm m}$02$^{\rm s}$.9 and
$\delta\,=$ 45$^\circ$ $02^\prime 26.0^{\prime\prime}$ that lies between
these two sources as the cluster center. 
This location was determined by estimating centers of ellipses 
at various surface brightness levels.

Another strong point source, called FS2 in the following, 
is found 9.2\arcmin~from the cluster core toward its northwest. 
This second source had already been detected in ROSAT data by 
Leahy \& Yin (2000). 
The location, X-ray flux, and information about radio counterparts of 
FS1 and FS2 are given in Table \ref{fsdata}. 
Lacking optical spectra, we can not determine whether these two source 
belong to the cluster or not. The only other known radio galaxy belonging 
to the cluster, WEIN 048 (Nilsson et al.\ 2000), was not in the field 
of view of the two Chandra observations.

We searched for large scale irregularities of the ICM surface brightness 
by constructing models of the undisturbed ICM surface brightness and
subtracting these model from the surface brightness map. 
We used two models of the undisturbed ICM. 
The first model is a ``standard King model'' that is symmetric around 
the cluster center:
\begin{equation}
\Sigma(r)\,=\,\frac{\Sigma_0}
{\left[1\,+\,\left(r/r_{\rm c}\right)^2\right]^{3\beta-\frac{1}{2}}}
\end{equation}
with the surface brightness $\Sigma(r)$, the normalization constant $\Sigma_0$,
the distance from the cluster center $r$, the cluster core radius $r_{\rm c}$, 
and the exponent $\beta$. 
using the fiducial value $\beta\,=$ 2/3 we fit a core radius of $r_c\,=$ 7.1\arcmin 
with a $\chi^2$-value of 317 for 41 degrees of freedom. 
Fitting also the $\beta$-parameter gives $r_c\,=$ 7.3\arcmin and $\beta\,=$ 0.68, and does 
not reduce the $\chi^2$-value substantially: 315 for 40 degrees of freedom. 
The surface brightness distributions after subtraction of this King model will 
show that the ICM distribution is quite irregular; we thus would not expect
that a symmetric King model gives a good fit.

The second model allows for variations of the ICM profile with polar angle.
This model assimilates some of the large scale irregularities of the 
ICM and allows us to search for irregularities on smaller angular scales.
Dividing the cluster into 18 pie sections of 20$^\circ$ opening angle each,
18 King models are fitted to the surface brightness. 
As shown in Fig.~\ref{fking} for two example pie sections, 
the fiducial value of $\beta\,=$ 2/3 yields unsatisfactory results: 
the observed surface brightness profiles show a more pronounced turn-down 
at cluster core distances between 5\arcmin~and 10\arcmin~than 
expected for $\beta\,=$ 2/3. 
Leaving $\beta$ as a free parameter, yields more satisfactory fits.
For polar angle values $\phi$ that are odd integer multiples of 10$^\circ$,
the surface brightness model is defined by the fitted King profiles. 
For intermediate values, we interpolate between the predicted surface
brightness of the adjacent two King models. To avoid modification of
the model by the radio galaxy 3C~129, we do not use the King model of
the corresponding pie section (polar angle 240$^\circ$-260$^\circ$), 
but interpolate between the King models of the two adjacent pie sections. 
We use these surface brightness models only to {\it find} excess emission; 
the statistical significance of excess features on small angular scales 
is rigorously evaluated based on count statistics as described above.

The residual surface brightness distributions smoothed with a Gaussian
of 18.8\arcsec FWHM are shown in Fig.\ \ref{smooth}, center and lower panels. 
The map derived from subtracting the symmetric King model emphasizes the
east-west elongation of the ICM: two regions with excess emission can be 
recognized: one is located at and toward the east of the cluster center, 
extending over at least 7\arcmin. The second one is located west/south-west 
of the cluster center. Note that the relative strength of these two excess
regions depends strongly on the choice of the location at which we 
center the symmetric King profile. Moving this location by 1.3\arcmin 
toward the west of the cluster, reduces the western excess substantially, 
especially the bright region around the head of the radio galaxy 3C 129.
Based on this surface brightness distribution, it is tempting to speculate
that the excess region near the cluster center shows remnants of a 
sub-cluster passage through the cluster center. 
The curvature of the 3C 129 radio tail may be caused by a sub-cluster that 
attracts the radio galaxy and whose dark matter center of gravity even lies 
in the south-west of the south-western excess region.
Upcoming observations of 3C 129 with the XMM-Newton satellite will shed more
light on the nature of these excess regions.
The map obtained by subtracting the 18-slice pie-model from the surface brightness
distribution does not show evidence for very sharp surface brightness gradients from 
large scale structure shocks or adjacent ICM phases of widely different 
X-ray emissivities.  
\section{X-Ray Surface Brightness Near the Two Radio Galaxies}
\label{features}
\subsection*{X-Ray Emission from 3C 129.1}
While the aimpoint of the ACIS~I observation is located near 3C~129.1, 
that of the ACIS~S observation lies near the head of 3C~129 at about 12\arcmin~from 
3C~129.1. Thus, although the latter observation has a three times longer 
integration time, its angular resolution is substantially lower: 
0.84\arcsec compared to 9\arcsec (FWHM) at the aimpoint of the observation 
and at 12\arcmin~off-axis, respectively. We therefore use only the ACIS I observation for 
the study of localized excess emission in the surrounding of 3C~129.1. 

The X-ray emission near 3C~129.1 is shown in Fig.\ \ref{rg1} together with
a contour plot of the high resolution radio data.
The high resolution radio map reveals a well defined northern jet, 
suggesting that the northern side approaches the observer.
An excess of X-ray emission over that from the ICM can be recognized up to 10\arcsec~from
the radio core. The excess emission corresponds to a luminosity of 
1.5$\times 10^{41}$ erg s$^{-1}$.
There is clear evidence that at least some of this emission comes from localized
sources rather than from diffuse gas. At about 4\arcsec~to the north of the radio 
core (source ``C'') we detect a total 7 counts (6 counts) from a sky region of 
1.1\arcsec 
(0.4\arcsec) radius , where 0.25 counts (0.04 counts) had been expected based on 
an annular comparison region of 11\arcsec inner radius and 20\arcsec outer radius.
In Table \ref{center} we list the locations, statistical chance probabilities, and 
fluxes for the extended ``center-source'' as well as for 5 potential point sources.
Using the Galactic neutral Hydrogen column density, we estimate a flux of
6.9$\times 10^{40}$  erg s$^{-1}$ for the center-source and fluxes
between 8.7$\times 10^{39}$ and 2.1$\times 10^{40}$  erg s$^{-1}$ for the point 
source candidates. 
The 0.5-5 keV energy spectrum of the combined excess X-ray photons 
can be described by a power law of spectral index $\alpha\,=$ 1.7$^{+0.4}_{-0.3}$
(with $E\,dN/dE\,\propto\,E^{-\alpha}$).
Using only the events from the point source candidates, the best-fit spectral index
is $\alpha\,=$ 1.3$^{+0.6}_{-0.4}$. 
Comparing the mean source spectrum luminosity with that of resolved sources in other 
X-ray faint galaxies \cite{Sara:00,Blan:01a,Irwi:02}, the luminosity of the 3C~129.1 
sources seems to be relatively high, while their energy spectrum is relatively soft.
This is especially true for source C which has the highest statistical significance,
the highest flux and a very soft energy spectrum (see Table \ref{center}).
\subsection*{X-Ray Surface Brightness Modification Near 3C 129}
\begin{figure}
\begin{center}
\epsfig{file=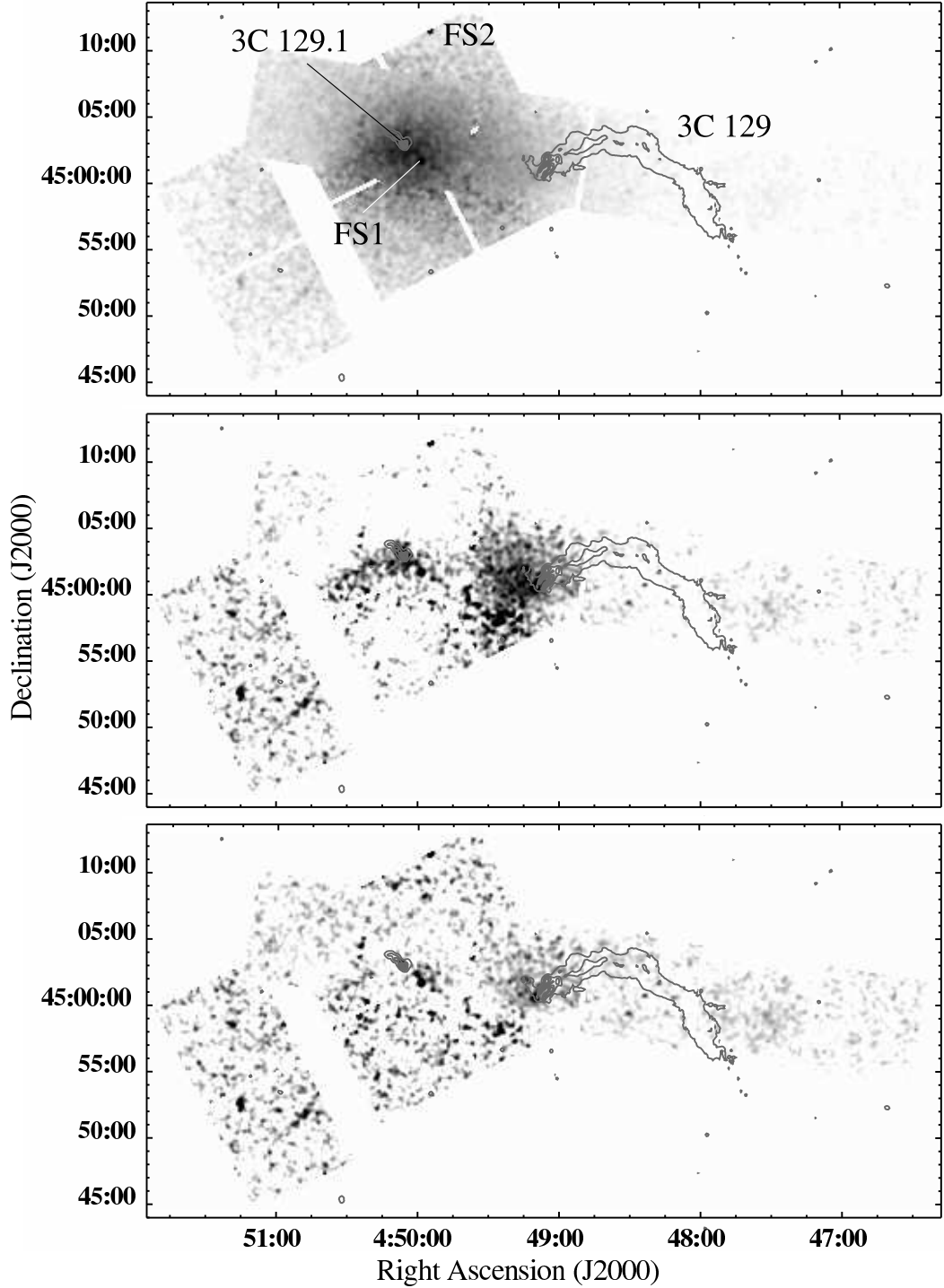,width=5.5in}
\end{center}
\caption{The gray-scale plot in the upper panel shows the flat-fielded Chandra 
0.5-5~keV data (ACIS I and ACIS S), smoothed with a Gaussian of 
18.8\arcsec FWHM. Segments of ACIS-I chip gaps are visible as white stripes 
outside the regions covered by the ACIS-S chips. The center panel shows the 
residual surface brightness distribution after subtraction of the symmetric 
ICM model; the lower panel shows the residual surface brightness 
after subtraction of the 18-slice pie model.
In all three panels the gray-scales are linear and high emission regions 
are darker than low emission regions; 
the surface brightness in the upper panel peaks at the location of 
FS 1 with 3.8$\times$10$^{-6}$ photons cm$^{-2}$ s$^{-1}$ arcsec$^{-2}$, 
at the radio core of 3C 129 it is 
1.0$\times$10$^{-6}$ photons cm$^{-2}$ s$^{-1}$ arcsec$^{-2}$; 
the gray of the ``noise'' at the very end of the 3C 129 tail is 
10$^{-7}$ photons cm$^{-2}$ s$^{-1}$ arcsec$^{-2}$; 
The center and lower panels use the same gray-scales with 
values of 2.1$\times$10$^{-6}$ photons cm$^{-2}$ s$^{-1}$ arcsec$^{-2}$
at the location of FS 1, and 10$^{-7}$ photons cm$^{-2}$ s$^{-1}$ arcsec$^{-2}$ 
at the noisy gray areas near the end of the tail of 3C 129, respectively.
The radio contours have been derived from the 330~MHz radio map and indicate the 
location of the radio galaxy 3C~129.1 at the cluster center and 3C~129 to the 
east of the cluster center. Contours are drawn at 0.0025, 0.035, 0.08, 0.12, 0.15, and
0.20 Jy beam$^{-1}$ (beamwidth 8\arcsec FWHM).}
\label{smooth}
\end{figure}

\begin{figure}
\begin{center}
\hspace*{-0.1cm} \epsfig{file=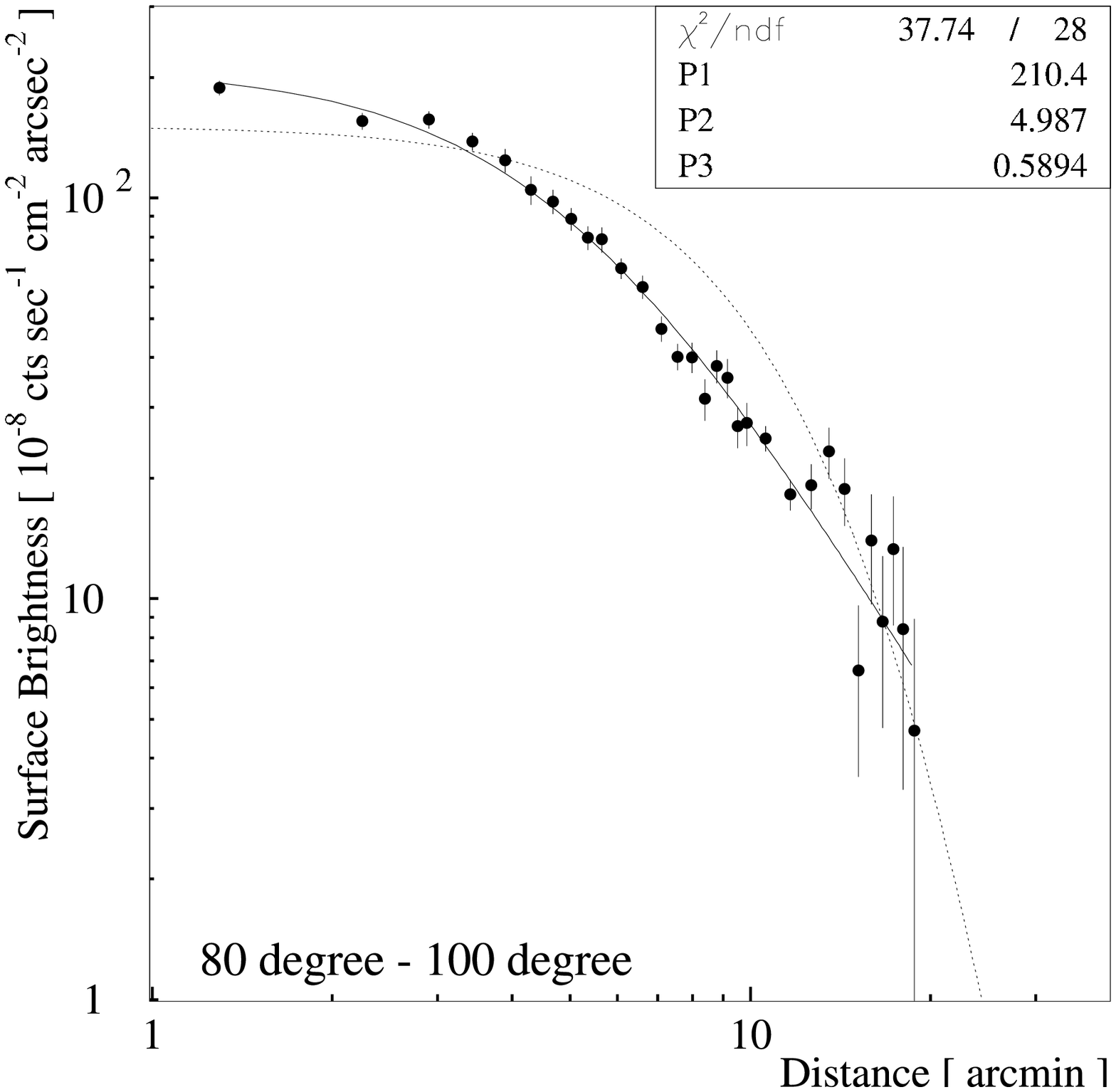,width=3.2in}\hspace*{0.2in}
\hspace*{-0.1cm} \epsfig{file=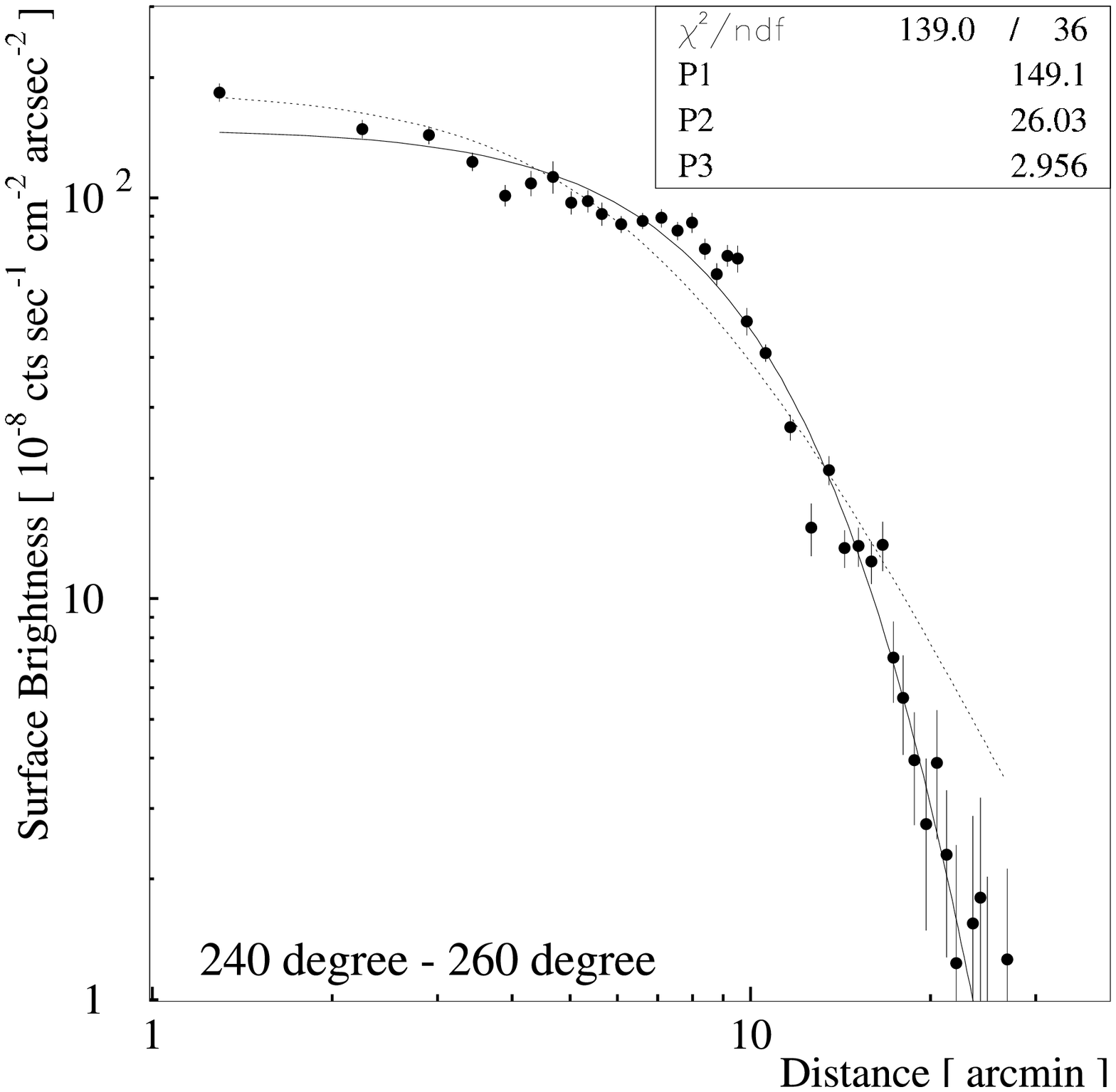,width=3.2in}
\end{center}
\caption{Radial 0.5-5~keV surface brightness distributions. The left
and right profiles have been derived from ``pie-sections'' of the
Chandra data with polar angles from 80$^\circ$ to 100$^\circ$ and from
240$^\circ$ to 260$^\circ$, respectively.  The full lines show fits of
King models and the best-fit values of the normalization factor
$\Sigma_0$ (in 10$^{-8}$ photons cm$^{-2}$ s$^{-1}$ arcsec$^{-2}$), 
core radius $r_{\rm c}$ (in arcmin), and $\beta$ are given in the upper 
right corner of each panel as P1, P2, and P3, respectively. 
In the left hand panel, the dotted line shows
the fitted profile of the right hand panel, and emphasizes the
difference of the two profiles. 
In the right hand panel, the dotted line shows a King model 
with $\beta\,=$ 2/3.}
\label{fking}
\end{figure}

\begin{figure}
\begin{center}
\epsfig{file=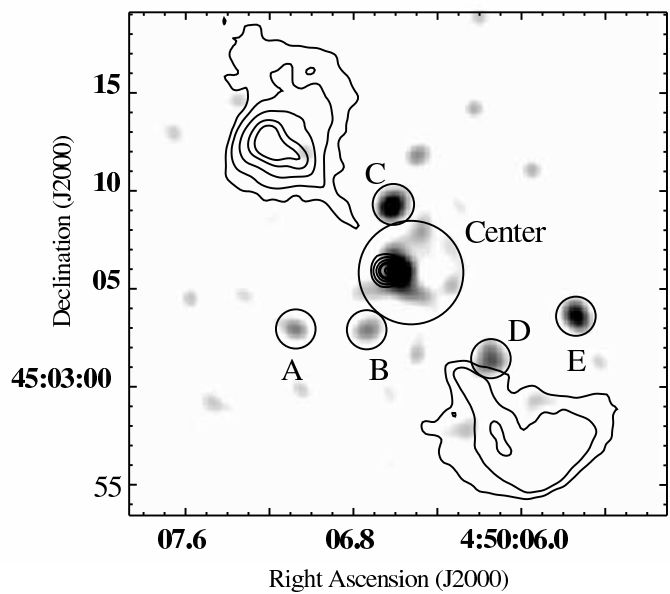,width=4.2in}
\end{center}
\caption{The gray-scale plot shows the 0.5-5~keV X-ray surface brightness 
at the location of the radio galaxy 3C~129.1, smoothed with a Gaussian of 
1.1\arcsec FWHM. The gray-scale is linear and peaks near the radio core of 3C 129.1 with
2.5$\times$10$^{-7}$ photons cm$^{-2}$ s$^{-1}$ arcsec$^{-2}$; 
for source A the surface brightness peaks at 1$\times$10$^{-7}$ photons cm$^{-2}$ s$^{-1}$ 
arcsec$^{-2}$.
The radio contours have been derived from the 8~GHz map 
with a beamwidth of 0.81\arcsec $\times$ 0.76\arcsec~FWHM with 9 logarithmically spaced levels 
from 2 to 512 mJy beam$^{-1}$. The high resolution radio map reveals a well
defined northern jet, suggesting that the northern side approaches the observer.
The photon excesses discussed in the text have been marked by circles.
}
\label{rg1}
\end{figure}

\begin{figure}
\begin{center}
\hspace*{-0.1cm} \epsfig{file=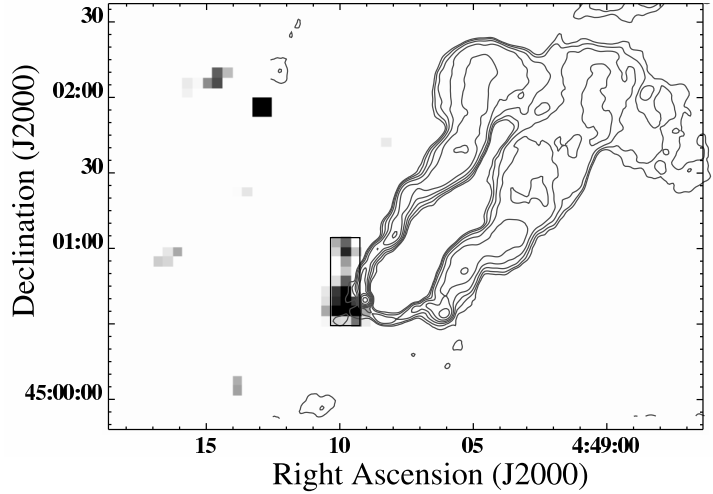,width=4.2in}
\end{center}
\caption{The gray-scale plot shows the residual 0.5-5~keV surface
brightness distribution near the head of the radio galaxy 3C~129 after
subtraction of the 18-slice pie model, smoothed with a 9.4\arcsec FWHM Gaussian.
The gray-scale is linear and peaks near the radio core of 3C 129 with
a value of 7.4$\times$10$^{-7}$ photons cm$^{-2}$ s$^{-1}$ arcsec$^{-2}$; 
the two bins at the bottom left side is a similar gray-tone have a surface 
brightness of 5.7$\times$10$^{-7}$ photons cm$^{-2}$ s$^{-1}$ arcsec$^{-2}$.
The radio contours have been derived from the 5~GHz VLA data 
with 1.8\arcsec FWHM beamwidth and the contours are logarithmically spaced from
1 to 32 mJy beam$^{-1}$. An X-ray bright region southeast of the radio core
can be recognized. The rectangle shows the region of the sky that we
used to determine the statistical significance of the excess.
}
\label{arm1}
\end{figure}

%
% FEATURES
%
The evidence for X-ray emission from the 3C~129 radio core and jet has
been described in Paper II. The excess emission was well localized and came from 
a sky region a couple of arc-seconds in diameter, located near the 3C~129 radio core. 
In Figs.\ \ref{smooth} and \ref{arm1}, we excluded this core and jet emission
from the analysis, by setting the residual surface brightness of these
sky regions to zero.  
The residual X-ray map of Fig.\ \ref{smooth} (lower panel) and the enlarged view of 
the sky region near the radio head of 3C~129 in Fig.\ \ref{arm1} 
reveal some evidence for an excess from a 
12\arcsec$\times$36\arcsec (6 kpc $\times$ 17 kpc) large region
at about 30\arcsec~(14 kpc) toward the east of the radio core. 
Using an annulus of 45\arcsec inner radius and 90\arcsec outer radius 
centered on the radio core as comparison region, we estimate the
statistical significance of the excess to be 3.7 $\sigma$
(ON: 131 counts, OFF: 92 counts).

%
% CAVITIES
%
The residual surface brightness map of Fig.\ \ref{smooth} (lower panel) 
does not show any evidence for excess or deficit emission along the tail of the radio galaxy 3C~129.
Indeed, in absence of strongly emitting sheets of compressed plasma,
we do not expect to detect any excess or deficit, given the small volume of the tail. 
Using a rectangular test region with a solid angle coverage of 
1.5\arcmin~$\times$ 6\arcmin~we can determine the ICM surface brightness to a
1$\sigma$ accuracy of between 4\% and 15\% near the radio head and near the extreme 
end of the radio tail, respectively. 
Assuming that (i) the tail is fully depleted of ICM, (ii) it lies in the plane 
of the sky, (iii) its depth along the line of sight equals its width 
perpendicular to the major tail extension, and (iv) that the ICM can be described 
by a King model with $r_{\rm c}$ = 7.1\arcmin and $\beta\,=$ 2/3 up to a cluster core
distance of 40', we expect a reduction of the ICM surface brightness 
of only between 3\% and 4\% for locations near the radio head and near the extreme 
end of the radio tail, respectively. 
Although $r_{\rm c}$ depends strongly on the location 
chosen as cluster center, the inferred reductions in the ICM surface brightness do not.
\section{Pressure Balance Along the Radio Tail of 3C~129}
\label{pb}
In this section we compare the thermal pressure of the ISM and ICM with the 
minimum non-thermal pressure of the 3C~129 radio tail.
The energy density of the synchrotron plasma (including magnetic field energy)
is minimized by the magnetic field (see Pacholczyk 1970):
\begin{equation}
B_{\rm min}\,=\,\sum_i~(6\,\pi)^{2/7}\,(1+k)^{2/7}\,\phi^{-2/7}\,V^{-2/7}\,\,
 \left( \sum_i c_{\rm 12,i}~ L_{\rm s,i}\right)^{2/7}
\label{min1}
\end{equation}
The $k$-factor is the contribution to the non-thermal pressure from  
relativistic protons over and above that from the relativistic electrons, 
$\phi$ is the volume filling factor for the radio emitting plasma, 
and $V$ is the emitting volume, well constrained by the observations. 
The sum runs over the considered frequency intervals $i$ of the observed radio 
spectrum, the value of $c_{12,i}$ depends weakly on the synchrotron spectral index and the 
lower and higher cutoffs of the $i^{\rm th}$ frequency interval, and $L_{\rm 
s,i}$ is the corresponding synchrotron luminosity.
We compute true minimum energy estimates by assuming a 
magnetic field of $B_{\rm min}$, a volume filling factor of $\phi\,=$ 1, 
no pressure contribution from electrons outside the energy range probed by the 
radio data, and no pressure contribution from relativistic protons ($k\,=$ 0). 
The finite angular resolution of the radio observations might overestimate the 
emission volume near the cores of the two radio galaxies 
and thus further lower the pressure estimate.
The non-thermal particle energy density is given by:
\begin{equation}
E_{\rm p}\,=\,\frac{4}{3}\,\frac{B_{\rm min}^2}{8\,\pi}
\label{min2}
\end{equation}
and the equation for the minimum isotropic pressure of the synchrotron plasma 
(including magnetic field pressure) reads:
\begin{equation}
P_{\rm min}\,=\,\frac{1}{3}\,E_{\rm p}\,+\,\frac{2}{3}\,\frac{B_{\rm 
min}^2}{8\,\pi}
\label{min3}
\end{equation}
In the latter equation we conservatively assumed a tangled magnetic field.

Based on these standard assumptions, $P_{\rm min}$ can be computed from 
the radio data alone. As a starting point we use either the 0.33~GHz, the 
5~GHz, or the 8~GHz surface brightness radio map. 
For each point along the radio tail we compute $P_{\rm min}$ using thin cylindrical 
segments, in other words: disks, seen edge on. The disk-axes are aligned with 
the radio tail. The width (and depth) of a disk is given by the two points at which the 
radio intensity drops to 10\% of its peak value.
For each disk we compute the mean radio flux $S_0$ at the frequency $\nu_0$ 
of the considered map, by averaging the surface brightness over the 
cross-section area of the disk. 
Given $S_0$, we construct the 0.33-8~GHz radio energy spectrum,
assuming the measured 0.33-1.4~GHz and 5-8 GHz spectral indices 
over the frequency ranges from 0.33-2.65 GHz and 2.65-8~GHz, respectively.
We use the ICM pressure from the spectroscopic ICM deprojection analysis
of the Chandra data \cite{Kraw:02} and neglect the pressure contribution 
of the ICM magnetic field.
Taylor et al.\ (2001) estimate that the ICM magnetic field might be as strong as 10 $\mu$G.
If this is true, the thermal and magnetic ICM pressures would be comparable at and near 
the head of the radio galaxy 3C 129.

\begin{figure}
\begin{center}
\epsfig{file=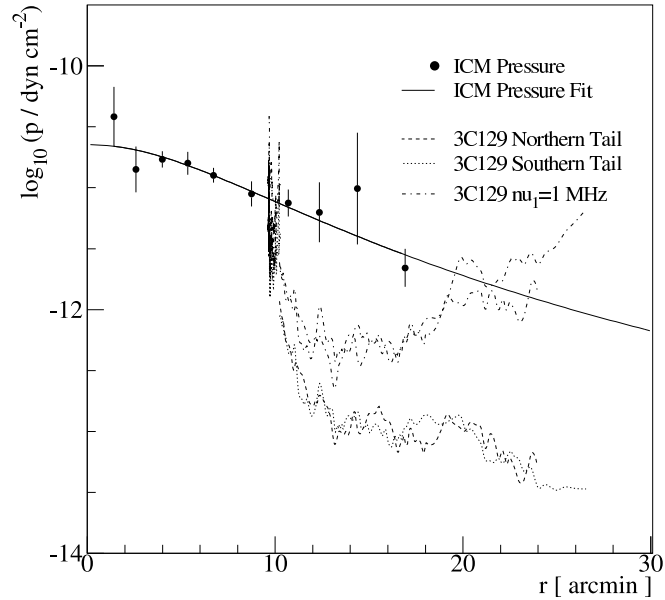,width=3.4in}
\end{center}
\caption{
The thermal ICM and non-thermal radio plasma pressures for 
the radio galaxy 3C~129 as function of distance to the cluster center
at $\alpha\,=$ 04$^{\rm h}$50$^{\rm m}$02$^{\rm s}$.9 and
$\delta\,=$ 45$^\circ$ $02^\prime26.0^{\prime\prime}$ (J2000).
The thermal ICM pressure from the spectroscopic deprojection
analysis of the X-ray data (Paper I) is shown by the solid line
and the data points. The dashed and dotted lines show the minimum 
pressure of the radio plasma for the northern and southern radio tail, respectively, 
computed for a minimum radio frequency $\nu_1\,=$ 330 MHz.
An estimate of the pressure contribution from very low energy electrons 
is shown by the dashed dotted lines, which have been computed under the
assumption that the radio spectrum extends down to $\nu_1\,=$ 1 MHz 
with the same spectral index as measured for 330 MHz - 1.4 GHz.}
\label{fpb}
\end{figure}

The pressure profiles are shown in Fig.\ \ref{fpb}. 
Over the first 13\arcsec (6.1 kpc) from the nucleus of 3C 129
the minimum pressure of the radio plasma 
exceeds the pressure of the ICM by a factor of up to $\simeq$5, 
suggesting either self-confinement, or, confinement by the 
cold disk-like remnant of the ISM that has been stripped off the galaxy 
by the relative motion of galaxy and cluster gas \cite{Toni:01}.
At 2\arcmin~(57 kpc) from the radio core, the minimum pressure fell by more than one
order of magnitude and shows strong wiggles, suggesting ``stormy local ICM weather\arcsec. 
Unfortunately, low X-ray photon statistics do not allow us to determine the ICM properties 
with sufficient statistical accuracy to assess the imprint of this stormy weather in the 
X-ray surface brightness or X-ray energy spectra.
Further away than 2\arcmin from the radio core, 
the minimum pressure levels off at 
roughly 1/40th of the ICM pressure.

The equipartition magnetic fields lie between 16 $\mu$G and 1 $\mu$G
near the beginning and the end of the tail, respectively. 
The lower frequency cutoff of 330 MHz corresponds to Lorentz factors of 
$\gamma_1$ $\simeq$ 4000 $(B/5~\mu\rm G)^{-1/2}$.
We estimate the possible pressure contribution of electrons with very low Lorentz factors
by extrapolating the radio energy spectrum down to $\nu_1\,=$ 1 MHz, 
corresponding to Lorentz factors of $\gamma_1$ $\simeq$ 220 $(B/5~\mu\rm G)^{-1/2}$.
Using the lower frequency cutoff, the inferred equipartition magnetic field strength 
increases to between 22 $\mu$G and 3 $\mu$G near the beginning and the end of the 
tail, respectively.
The corresponding pressure estimates are shown by the dashed dotted lines in Fig.\ \ref{fpb}.

For the first 10\arcmin~of the tail, the radio spectrum is sufficiently flat so
that the lower cutoff frequency does not qualitatively alter the result:
the minimum non-thermal pressure lies substantially below the ICM pressure.
We will discuss the implications in the next section.
\section{Summary and Discussion}
\label{disc}
%
% the cluster center and 3C 129.1
%
The Chandra observations show that the radio galaxy 3C 129.1 is associated with
an extended X-ray excess. The excess can be divided into 5 point sources and a central
component. On the small angular scale of the central component (4\arcsec diameter) 
the ISM is not expected to contribute to the excess. 
We therefore consider it most probable that the central excess emission is  
produced by galactic sources and possibly also by the central AGN.

%
% 3C 129
%
For the radio galaxy 3C~129 we find, in addition to the core and jet emission described in Paper~II,
some evidence for extended X-ray emission from a 
12\arcsec$\times$36\arcsec (6 kpc $\times$ 17 kpc) large region to the east 
of the radio core. If confirmed with higher statistical significance,
a natural explanation of the excess would be thermal 
emission from plasma compressed and heated by the bow shock that
precedes the galaxy. 

Comparing the minimum pressure of 3C 129 radio tail with the thermal 
ICM pressure, the radio plasma appears to be under-pressured by 
a factor of between 10 and 40. Over the first 10\arcmin of the radio tail, 
the radio spectrum is sufficiently flat so that the results depend only weakly on 
the assumed minimum Lorentz factor of the non-thermal electron distribution.

Remarkably, a very similar pressure mismatch has been found for the kilo-parsec 
radio features of a number of FR~I and FR~II radio galaxies 
(see Hardcastle \& Worral 2000 and references therein).
For most radio galaxies the magnetic field inside the radio plasma is 
poorly constrained toward higher values and might dominate the plasma pressure.
In the case of 3C 129, we have an additional handle on the magnetic field strength from
the constraints on the radiative age of the radio-electrons:
If the missing pressure of the radio plasma is provided by the magnetic 
field alone, the implied field strengths of between 15$\mu$G and 30$\mu$G 
translate into very high lower limits on the galaxy velocity of between 
11,000 km s$^{-1}$ and 31,000 km s$^{-1}$ (see Eq.\ \ref{vgal}).
Buoyant motion of the radio plasma would proceed with a velocity 
lower or equal to the
ICM sound speed, and would thus not qualitatively alter our conclusions.
The apparent pressure mismatch thus suggests one of the following possibilities:
(i) thermal or relativistic protons and/or thermal or relativistic
low-energy electrons carry the major fraction of the radio plasma's pressure;
(ii) the volume filling factor of the radio plasma $\phi$ is substantially 
smaller than unity: $\phi\,=$ 0.0016 - 0.018; 
(iii) the $B$-field is much larger than the
equipartition magnetic field and electrons are re-accelerated 
inside the radio tail;
(iv) the radio galaxy lies behind or in front of the plane of the cluster 
so that the true ambient ICM pressure is lower than assumed in our analysis.
The detection of an ``ICM cavity'' associated with the radio tail would make 
it possible to estimate the volume filling factor. We hope to detect such a 
cavity with an upcoming XMM-Newton observation of 3C 129.
\hspace*{2cm}\\[2ex]
{\bf Acknowledgements}
We thank Lewis Knee and Charles Kerton for advice and assistance with
reduction of the DRAO HI data.  The Canadian Galactic Plane
Survey is a Canadian project with international partners, and is
supported by the Natural Sciences and Engineering Research Council
of Canada. The Dominion Radio Astrophysical Observatory is a
national facility operated by the National Research Council of
Canada. We thank G. Taylor for providing FITS images of his radio maps and
for higher resolution versions of the 8 GHz data than those used in
his paper.  We are grateful to the anonymous referee for very valuable comments.
The analyzes in this paper were partially supported by
NASA grants GO1-2135B (HK), GO1-2135A (DEH), and by NASA contract
NAS8-39073 (DEH).  WL is a National Research Council Postdoctoral
Fellow.  Basic research in astronomy at the Naval Research Laboratory
is funded by the Office of Naval Research.

\newpage
\onecolumn
%
% Data Sets
%
\begin{table}
 \caption{3C 129 Chandra Data Sets.}
 \label{data1}
 \begin{tabular}{@{}p{3.5cm}p{1.5cm}p{2.5cm}p{2cm}p{2.5cm}} \hline
Description & Energy & Integration Time & Ang.~Res. (FWHM)& Reference\\
\hline
Chandra ACIS I &0.5-8~keV& 9.3 ksec &  0.84\arcsec (aimpoint) & Krawczynski 2002\\
Chandra ACIS S &0.5-8~keV&28.3 ksec &  0.84\arcsec (aimpoint) & Krawczynski 2002\\ 
\hline
 \medskip
\end{tabular}
\end{table}
\begin{table}
 \caption{Radio Data Sets Used in this Paper.}
 \label{data2}
 \begin{tabular}{@{}p{3cm}p{1.3cm}p{1.3cm}p{1.3cm}p{1.5cm}p{1.3cm}p{2.5cm}} \hline
Frequency  & Instrument & New Data & 
Ang. Res.\ (FWHM) & Noise RMS $\rm \left[mJy beam^{-1}\right]$ & DR & Reference\\
\hline
HI                 & DRAO & Y &    1\arcmin & ---   & ---  & this work \\ 
330 MHz            & VLA  & N &    8\arcsec & 0.75  & 400  & Lane et al.\ 2002\\
1.4 GHz            & VLA  & Y &   16\arcsec & 0.1   & 3000 & this work\\ 
4.7 GHz (3C 129)   & VLA  & N &  1.8\arcsec & 0.072 & 545  & Taylor et al.\ 2002\\
8   GHz (3C 129)   & VLA  & N &  1.8\arcsec & 0.04  & 925  & Taylor et al.\ 2002\\
8   GHz (3C 129)   & VLA  & N & 0.83\arcsec & 0.03  & 1287 & Harris et al.\ 2002\\
8   GHz (3C 129.1) & VLA  & N &  0.8\arcsec & 0.012 & 225  & Taylor et al.\ 2002\\
\hline
 \medskip
\end{tabular}
\end{table}
%
% Field Sources
%
\begin{table}
 \caption{X-ray sources not associated with known cluster galaxies.}
 \label{fsdata}
 \begin{tabular}{@{}p{2cm}p{2cm}p{2cm}p{2.5cm}p{6cm}} \hline
Name & RA (Eq.\ 2000) & Decl (Eq.\ 2000) & 0.5-5 keV Flux$^a$ & 
Radio Counterpart \\
&$\left[\mbox{hh:mm:ss.ss}\right]$  &
$\left[\mbox{dd:mm:ss.ss}\right]$  & 
$\left[\mbox{erg\,cm}^{-2}\,\mbox{s}^{-1}\right]$  & 
\\ \hline
FS1 & 04:49:58.5 & 45:01:44.8 &  5.13$\times$10$^{-14}$ & 0.24 mJy @4.8 GHz, 
0.11 mJy @8 GHz\\
FS2 & 04:49:54.8 & 45:11:31.4 &  8.85$\times$10$^{-14}$ & no \\
\hline  
\end{tabular}
\hspace*{1cm}\\
$^a$ Observed flux.\\
\end{table}
\begin{table}
 \caption{X-ray Sources of 3C~129.1}
 \label{center}
 \begin{tabular}{@{}p{2cm}p{2cm}p{2cm}ccp{1.5cm}p{1.5cm}p{2cm}p{1.7cm}} \hline
Name & RA (Eq.\ 2000) & Decl (Eq.\ 2000) & $N_{\rm ON}$ & $\eta$ $N_{\rm OFF}^{\,a}$ & 
Chance& $\varepsilon_{\rm 0.5-5 keV}$$^b$ & Flux$^c$ & Luminosity $^d$\\
&
$\left[\mbox{hh:mm:ss.ss}\right]$  &
$\left[\mbox{dd:mm:ss.ss}\right]$  & & & Probab.&
$\left[\mbox{keV}\right]$ &
$\left[\mbox{erg\,cm}^{-2}\,\mbox{s}^{-1}\right]$ &
$\left[\mbox{erg~s}^{-1}\right]$ \\ \hline
A       & 04:50:07.075&45:03:02.91& 3 & 0.40& 8.0$\times 10^{-3}$ & 2.17$\pm$0.29
	& 2.0 10$^{-15}$ & 8.1$\times 10^{39}$\\ 
B       & 04:50:06.729&45:03:02.92& 4 & 0.40& 7.9$\times 10^{-4}$ & 3.58$\pm$0.68
	& 2.7 10$^{-15}$ & 1.1$\times 10^{40}$\\
C       & 04:50:06.618&45:03:09.18& 7 & 0.40& 2.3$\times 10^{-7}$ & 2.27$\pm$0.17
	& 5.0 10$^{-15}$ & 2.1$\times 10^{40}$\\
Center  & 04:50:06.584&45:03:05.92& 25& 2.80& 6.7$\times 10^{-16}$& 2.48$\pm$0.15
	& 1.7 10$^{-14}$ & 7.0$\times 10^{40}$\\
D       & 04:50:06.149&45:03:01.35& 5 & 0.40& 6.2$\times 10^{-5}$ & 3.69$\pm$0.58
	& 3.5 10$^{-15}$ & 1.5$\times 10^{40}$\\
E       & 04:50:05.740&45:03:03.61& 6 & 0.40& 4.1$\times 10^{-6}$ & 4.16$\pm$0.46
	& 4.2 10$^{-15}$ & 1.8$\times 10^{40}$\\ 
\end{tabular}
\hspace*{1cm}\\
$^a$ Expected number of ``background'' counts in the ON region.\\
$^b$ Mean energy of excess 0.5-5 keV Chandra photons.\\
$^c$ Observed flux.\\
$^d$ ``Unabsorbed luminosity'', assumes N$_{\rm H}\,=$ 8.8$\times$$10^{21}$~cm$^{-2}$.\\
\end{table}
%
% Nh
%
%
% wendy
%
%
% ICM, residuals, 330 MHz map
%
%
% King profiles
%
%
% 3C129.1
%
%
% 3C129
%
%
% pressure balance
%
\end{document}